\documentclass[aps,twocolumn,showpacs,prl,superscriptaddress]{revtex4}
\usepackage{amsfonts}
\usepackage{upgreek, graphicx, subfigure, amsmath}
\usepackage{epstopdf}
\usepackage{color,soul,ulem}

\newcommand{\rr}{{\mathbf r}}

\newcommand{\ket}[1]{\left\vert #1 \right\rangle}

\newcommand{\sub}[1]{{\textrm{#1}}}

\begin{document}

\title{Quantum galvanometer by interfacing a vibrating nanowire and cold atoms}

\author{O.\ K\'{a}lm\'{a}n}
\affiliation{Research Institute for Solid State Physics and Optics of the Hungarian Academy of Sciences, H-1525 
Budapest P.O. Box 49, Hungary}
\author{T.\ Kiss}
\affiliation{Research Institute for Solid State Physics and Optics of the Hungarian Academy of Sciences, H-1525 
Budapest P.O. Box 49, Hungary}
\author{J.\ Fort\'{a}gh}
\affiliation{Physikalisches Institut, Universit\"{a}t T\"{u}bingen, Auf der 
Morgenstelle 14, 72076 T\"{u}bingen, Germany}
\author{P.\ Domokos}
\email{domokos@szfki.hu}
\affiliation{Research Institute for Solid State Physics and Optics of the Hungarian Academy of Sciences, H-1525 
Budapest P.O. Box 49, Hungary}


\begin{abstract}

We evaluate the coupling of a Bose-Einstein condensate of ultracold, paramagnetic atoms to the magnetic field of the current in a mechanically vibrating carbon nanotube within the frame of a  full quantum theory. We find that the interaction  is strong enough to sense quantum features of the nanowire current noise spectrum by means of hyperfine-state-selective atom counting. Such a non-destructive measurement  of the electric current via its magnetic field corresponds to the classical galvanometer scheme, extended to the quantum regime of charge transport. The calculated high sensitivity of the interaction in the nanowire-BEC hybrid systems opens up the possibility of quantum control, which may be further extended to include other relevant degrees of freedom.

\end{abstract}

\maketitle

Carbon nanotube (CNT) technology has evolved to the point, where CNTs can be produced with a variety of mechanical and electrical properties \cite{Harris09}. Besides applications in chemical \cite{Kong00}, biological \cite{Lin04}, and mass sensing with up to single atom resolution \cite{Chiu08}, high quality nanomechanical resonators \cite{Huttel09nl}, one-dimensional electric transport effects and their coupling to mechanical motion of the CNTs have been observed \cite{Steele09,Lassagne09}. The increasingly fine control of these degrees of freedom anticipates the manipulation of CNTs at a quantum mechanical level that has been recently achieved in other nanomechanical systems  \cite{OConnell2010}.

Atomic physics has undergone a breathtaking evolution since laser cooling and trapping techniques allowed for the preparation of  localized and isolated atom samples \cite{Nobel1997}. The thermal noise has been reduced to the ultimate quantum noise level where the ``ultracold atoms'' (below 1$\mu$K) form a degenerate quantum gas, called a Bose-Einstein condensate (BEC) \cite{castin2001}. This cloud of atoms can be controlled in all relevant degrees of freedom with an unprecedented precision. The cloud can be positioned on the submicrometer scale in magnetic \cite{fortagh07} or optical traps \cite{Grimm2000}.  The internal electron dynamics can be driven by external laser or microwave fields which allow for the precise preparation as well as the high-efficiency detection of the electronic state.  The system of trapped neutral atoms in the collective BEC state is an ideal probe of external fields \cite{Wildermuth2005}.

The first attempts to make a BEC interact with a CNT have been reported only recently. Atom scattering from the CNT's van der Waals  potential \cite{Giergling2011}, and the field ionization due to charged suspended nanotubes have been observed\cite{Goodsell2010Field}. Recent proposals have discussed how to make use of the CNT as a current-carrying thin nanowire to tighten the magnetic trapping potential for cold atoms \cite{Folman09} and how to form nanoscale plasmonic atom traps along silver decorated CNTs \cite{Murphy09}.   The integration of carbon nanotubes and atomic Bose-Einstein condensates opens up new avenues towards hybrid systems coupling these objects at a quantum level in a controlled way. One can envisage the coherent interfacing of very different degrees of freedom such as electronic, mechanical, and spin variables. The question is whether there is a suitable interaction between selected degrees of freedom and whether the  cross-coupling is strong enough  to design useful quantum devices. A variety of novel nanodevices for precision sensing, quantum measurement, and quantum information processing could be developed on the basis of CNT-BEC coupling.

In this Letter we theoretically evaluate the interaction between the current through the nanowire and atoms in a condensate. We describe how the internal atomic dynamics in the hyperfine states couples to the magnetic field generated by the CNT current. Starting from a fully quantum model we calculate the time evolution of the atomic system and construct a scheme to measure the current via an atomic observable. All this leads to the conclusion that quantum dynamical properties of the CNT are detectable by making use of the mature technology of ultracold atoms. A more general implication is that the other degrees of freedom of the CNT and BEC which take part in the dynamics can also be accessed and possibly manipulated in variants of this scheme.

For the sake of concreteness, we focus on the measurement of the current through the CNT in the  galvanometer scheme, i.e., when the electric current is sensed in a non-destructive way via its 
magnetic effect. We are interested in the quantum transport limit of a mesoscopic conductor. In principle, a quantum object such as a spin-1/2 particle precessing in the magnetic field is sensitive to the quantum properties of the current. The  full counting statistics of the  charge transport process \cite{Bednorz2010Quasiprobabilistic} can be reconstructed from the density matrix of the spin \cite{Levitov03}. This scheme is sometimes referred to as the \textit{quantum galvanometer} \cite{Levitov96}. However, its realization has to meet several conditions (stability, sensitivity and detectability, see below), which has seemed out of 
the question so far.  

In our scheme the fictitious spin of the quantum galvanometer is realized by a cloud of long-term trapped atoms. One can make use of the collective BEC state which greatly enhances the 
sensitivity of the internal hyperfine dynamics to external magnetic fields \cite{Vengalattore2007High}. Moreover, a direct readout method for the spin state is available since the  populations in the magnetic sublevels can be counted by state selective ionization with single-atom resolution \cite{fortagh_detect}.  The populations are expressed in the diagonal elements of the spin density matrix. In principle, the off-diagonal elements could also be measured, which is required to determine the   generating function of the full counting statistics. Here 
we propose a simpler measurement restricted to the diagonal elements, since it is already enough to characterize some of the quantum features of the current. In particular, we will show that the ordinarily defined current noise spectrum, 
\begin{equation}
\label{eq:noise_spectrum}
S(\omega) = \int_{-\infty}^{\infty} d\tau e^{i \omega \tau} 
\left<\hat{I}(0)\hat{I}\left(\tau\right)\right> \,,
\end{equation}
is directly related to the time evolution of the populations in the given magnetic sublevels. The proposed galvanometer allows for the measurement of the spectrum by scanning the adjustable variable $\omega$ which spans both the negative and positive frequency ranges.  Asymmetry of the measured spectrum around $\omega=0$ would reveal quantum features  \cite{NazarovBlanter}. This is due to the fact that asymmetry of $S(\omega)$ is equivalent with the non-commutativity of the current operator at different instances, as one can easily check in the above equation.

\begin{figure}[htb]
\includegraphics[width=0.86\columnwidth]{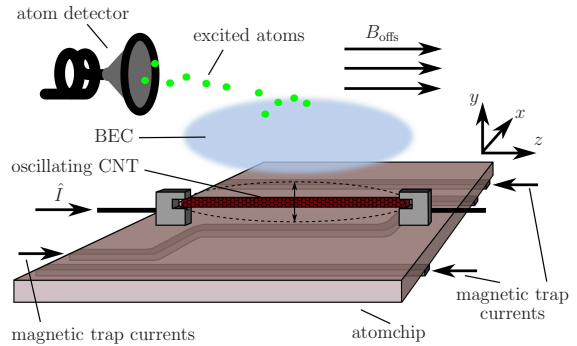}
\caption{The quantum galvanometer on an atomchip. A BEC is loaded into the  
magnetic microtrap created by the classical electric currents through  
integrated conductors on a dielectric substrate \cite{fortagh07} (represented 
on the bottom of the chip). A suspended carbon nanotube is also part of the 
electric circuit \cite{Folman09} and transports the quantum current. The 
oscillating CNT  creates a magnetic field in the 10 MHz range interacting with 
the hyperfine transitions of the atoms. Atoms transferred to untrapped states 
are detected by a single-atom detector \cite{fortagh_detect}.}
\label{fig:setup}
\end{figure}

The scheme is presented in \ref{fig:setup} for a possible architecture of building the galvanometer on an integrated platform, the so-called ``atomchip"  \cite{fortagh07,Folman_review}. It is a compact, substrate-based electric circuit of currents which create highly tunable magnetic traps for the neutral atoms. The chip may support a contacted CNT on the side facing the BEC. The suspended CNT is a high-Q mechanical oscillator \cite{Sazonova04,Witkamp06,Huttel09nl,OConnell2010}, which is driven to oscillate coherently with large amplitude. We note that in order to simplify our calculations, here we assume that the vibration of the CNT is driven by a mechanical source (e.g. a piezo crystal) instead of electric fields. Therefore the CNT current generates a harmonically time-varying magnetic field. In the neighboring BEC, this field may excite transitions between the hyperfine states. The key role of the mechanical vibration of the CNT consists in creating the resonance conditions such that the low-frequency components of the current noise spectrum be close to resonance with the hyperfine transitions (Zeeman splitting), c.f.\ Fig. 2.  Initializing the atoms in the
BEC in a well-defined hyperfine state, the number of atoms transferred to another state follows a statistics determined by the low-frequency part of the current noise spectrum. The noise source can be arbitrary, e.g., thermal or intrinsic quantum noise; for generality, we will represent the current as a quantum operator in the derivation. Finally, the measurement is accomplished by detecting the  number of transferred atoms in a given time period, which can be performed  by means of state-selective ionization technique at the single-atom resolution level \cite{fortagh_detect}.

To be specific, we consider ultracold ${}^{87}$Rb atoms in the hyperfine 
state $F=1$. The atomic spin $\hat{\mathbf F}$ interacts with the magnetic 
field according to the Zeeman term $H_{Z} = g_{F} \, \mu_{\mathrm B} 
\hat{\mathbf F}{\mathbf B}(\rr)$, where $\mu_{\mathrm B} = e\hbar/2 m_{e}$ is 
the Bohr magneton, the Land\'{e} factor is $g_{F}=-1/2$, and $\hat{\mathbf F}$ 
is measured in units of $\hbar$. The dominant component of the magnetic field is 
a homogeneous offset field $B_{\rm offs}$ in the $z$ direction.  
The eigenstates of the spin component $\hat F_{z}$ are then well separated by 
the Zeeman shift. These are the magnetic sublevels labelled by $m=-1,0,1$. On 
top of the offset field, there is an inhomogeneous term which creates an 
ellipsoidal potential around the minimum of the total magnetic field. The atoms 
are then subject to the static potential 
$V_{m}(\rr)= -m V_{\mathrm T}(\rr)+g_{F}\mu_{\mathrm B} m B_{\mathrm{offs}},$ 
where
$V_{\mathrm T}(\rr)=\frac{M}{2}\left[\omega_{r}^{2}\left(x^{2}+y^{2}\right)
+\omega_{z}^{2}z^{2}\right]$, with $M$ being the atomic mass. The potential is 
diagonal in the $\hat F_{z}$ basis, and is confining only for $m=-1$. 

\begin{figure}[thb]
\includegraphics[width=0.86\columnwidth]{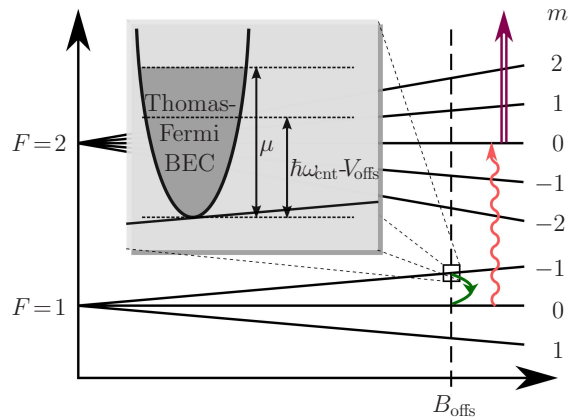}
\caption{Level scheme of the ${}^{87}$Rb atoms. The inset shows the resonance 
condition for the $m= -1 \leftrightarrow 0$ rf transition induced by the 
time-varying magnetic field component due to the oscillating CNT. The shaded 
region corresponds to the band of energies in the Thomas-Fermi approximation. 
The detection of atoms in the spatially separated state $F=1, m=0$ is performed 
by a resonant microwave excitation to $F=2, m=0$ (curly arrow), and then by a 
two-photon ionization process \cite{fortagh_detect} (double arrow).}
\label{fig:levelscheme}
\end{figure}

Let us consider a single carbon nanotube of length $L$ which is electronically 
contacted and carries a current $\hat I (t)$. It generates a magnetic field 
that interacts with the atomic spin. Similar coupling has been considered \cite{Treutlein07} 
between a vibrating nanomagnet and a BEC. The CNT is 
aligned with the $z$ axis (see \ref{fig:setup}) having a mean distance 
$y_{0}$ from the condensate. This distance is large enough ($y_0 \geq 1 \mu$m to 
avoid Van der Waals-type interactions between the atoms and the CNT 
\cite{Fermani07,Ristroph2005,Goodsell2010Field,Fortaghnew,Giergling2011}). The CNT is driven 
to mechanically oscillate harmonically at an angular frequency 
$\omega_{\rm cnt}$ in the 10 MHz range \cite{Sazonova04,Witkamp06} and with an 
amplitude $a$ in the $y-z$ plane ($a\ll y_{0}$). The resulting time-dependent 
magnetic field $\hat{\mathbf B}_\sub{cnt} (\rr, t)$ is an operator, since its 
source is the current operator $\hat I$. With a proper tuning of the Zeeman 
splitting via the offset $B_\sub{offs}$, the $x$ and $y$ components of 
$\hat{\mathbf B}_\sub{cnt}$ can quasi-resonantly generate transitions between 
the magnetic sublevels $m$. Such a transfer of trapped atoms into untrapped ones 
is the underlying mechanism of the rf outcoupler of an atom laser 
\cite{Steck98}. Here a fast detection of the spatially separated component $m=0$ 
will be required by means of  combined microwave transition and 
two-photon ionization process \cite{fortagh_detect} via the state $F=2, m=0$ 
(see \ref{fig:levelscheme}). 

\begin{figure}[tbh]
\includegraphics[width=0.86\columnwidth]{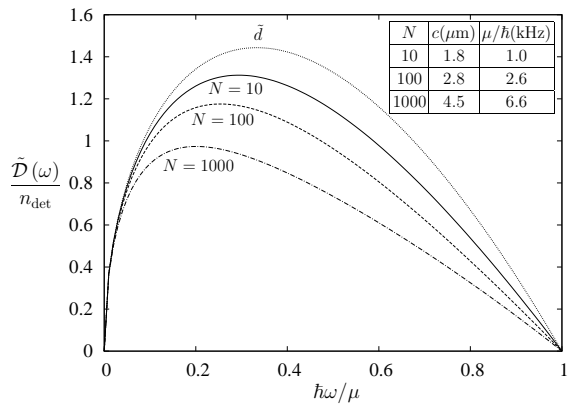}
\caption{The dotted curve shows the approximation $\tilde{d}\left(\omega\right)$ 
from \ref{eq:D_Omega}, compared to the exact 
$\tilde{\mathcal{D}}\left(\omega\right)$ in units of $n_{\rm det}$ for 
different number of trapped atoms $N$ for a trap with radial and axial 
frequencies $\omega_{r}=2\pi\!\times\!500\,\mathrm{s}^{-1}$, and 
$\omega_{z}=2\pi\!\times\!109\,\mathrm{s}^{-1}$, respectively. The corresponding 
longitudinal extent $c$ of the atom cloud and the chemical potential are 
presented in the table. Note that the radial confinement of the cloud is 
determined by $b/c=\omega_{z}/\omega_{r}$. Here we assumed the length of 
the CNT, and the CNT-BEC distance to be $L=2$ $\mu$m, and 
$y_{0}=4$ $\mu$m, respectively.}
\label{fig:d_omega}
\end{figure}

In the following, we will describe the BEC-CNT interaction based on a model Hamiltonian [\ref{eq:H_manyatom}] which treats all the relevant degrees of freedom quantized. 
Within the Thomas-Fermi approximation (large condensate limit, \ref{eq:Thomas_Fermi}) and to first order in perturbation theory [\ref{eq:time_ev}],  we will solve the dynamics for the time 
evolution of the atomic state populations [\ref{eq:N_Omega}]. Thereby, we will obtain a relation between the population in the externally monitored hyperfine state and the current noise spectrum in a closed form [\ref{eq:atomspectrum}]. 
This relation is a convolution which includes a detection spectral function characterizing the resolution of the measurement [\ref{eq:resolution_function}]. We will present its functional form in \ref{fig:d_omega} and discuss how its magnitude depends on the physical parameters of the setup.

The ultracold atom cloud can be described by the spinor field\cite{Ho1998} 
$\sum_{m} \hat \Psi_{m} \ket{F=1, m}$.  The very general many-body Hamiltonian is given by
\begin{multline}
\label{eq:H_manyatom}
\!{\mathcal H}= \!\int \! d^3\rr \sum_{m,m'=-1}^1 \hat \Psi_m^\dagger (\rr) 
\Biggl[  \left( -\frac{\hbar^2 \nabla^2}{2M} +  V_{m} (\rr)  \right) 
\delta_{m,m'} \\ 
+ \frac{g_{m,m'}}{2} \, \hat \Psi_{m'}^\dagger(\rr) \hat\Psi_m (\rr)  
 +  g_F \mu_{\mathrm B} \hat{\mathbf F}  \hat{\mathbf B}_{\rm cnt} (\rr) 
\Biggr]   \hat\Psi_{m'} (\rr),  
\end{multline}
where the first term is the kinetic energy. The atom-atom interaction in this 
ultracold regime is s-wave scattering, spin flipping collisions are negligible.  
In principle, the coupling term induces a dynamical back action on the current and the vibration of the CNT. Here we will disregard the back action, however, we note that it could be the source of interesting new effects in similar schemes. 

We assume that there is a condensate of a number of $N$ atoms in the trapped 
$m=-1$ magnetic sublevel.  For convenience, we separate the condensate part 
$\sqrt{N} \phi_\mathrm{BEC}$ from the excitations $\delta \hat \Psi_{m}$ which 
will be treated perturbatively. The condensate wavefunction is 
$\phi_\mathrm{BEC}(\rr ,t)=e^{-\frac{i}{\hbar}\mu\rq t}\phi_\mathrm{BEC}(\rr),$
where $\mu\rq$ is the chemical potential, and $\varphi_{\mathrm{BEC}}(\rr)$ obeys 
the Gross-Pitaevskii equation \cite{Steck98}.
We restrict ourselves to the Thomas-Fermi 
approximation, i.e., we neglect the kinetic energy term in the Gross-Pitaevskii 
equation. Then the condensate wavefunction is 
\begin{subequations}
\begin{align}
\phi_{\mathrm{BEC}}(\rr) &=
\sqrt{\frac{\mu-V_{\mathrm{T}}(\rr)}{Ng}}\,, \; \mbox{where}
\label{eq:Thomas_Fermi}
\\
\mu=\mu\rq-\frac{1}{2}\mu_{\mathrm{B}}B_{\mathrm{offs}}&=
\left(Ng\frac{15}{8\pi}\omega_{r}^{2}\omega_{z}\right)^{\frac{2}{5}}
\left(\frac{M}{2}\right)^{\frac{3}{5}}\,.
\end{align}
\end{subequations}
The condensate shape is an ellipsoid with principal semi-axes 
$b=\sqrt{\frac{2\mu}{M\omega_{r}^{2}}}$ and 
$c=\sqrt{\frac{2\mu}{M\omega_{z}^{2}}}$. The atom-atom collisions are accounted for by $g=4\pi\hbar^{2}a_{\mathrm{s}}/M$, where $a_{\mathrm{s}}$ is the 
$s$-wave scattering length ($a_{\mathrm{s}}$=5.4 nm for ${}^{87}$Rb). We 
will neglect the back action of the other spin components onto the condensate and assume that the condensate is intact during  the interaction time.

To first order in the perturbations, the equation of motion for the component 
$\delta\hat{\Psi}_{0}$, after some straightforward calculation, is obtained as
\begin{multline}
i\hbar\frac{\partial}{\partial t}\delta\hat{\Psi}_{0}(\rr,t)\!=\! 
\left(-\frac{\hbar^{2}\nabla^{2}}{2M}+
Ng\left|\phi_{\mathrm{BEC}}(\rr,t)\right|^{2}\right)
\delta\hat{\Psi}_{0}(\rr,t) \\ 
-\frac{\mu_{\mathrm{B}}}{2\sqrt{2}}\left(\hat{B}_{x}-i\hat{B}_{y}\right)
\left(\sqrt{N}\phi_{\mathrm{BEC}}(\rr,t)+\delta\hat{\Psi}_{-1}(\rr,t)\right), \label{eq:time_ev} 
\end{multline}
where we have omitted the subscript 'cnt' when writing the components of the 
magnetic field created by the current through the nanotube, which for 
$a\ll y_{0}$  can be approximated as
$\hat{B}_{i}(\rr,t)\approx \hat{B}_{i0}(\rr,t)+\delta 
\hat{B}_{i}(\rr,t)\cos\left(\omega_{\mathrm{cnt}}t\right),
\; \; (i=x,y)$.
The last term with the quantum field $\delta\hat{\Psi}_{-1}$ in 
\ref{eq:time_ev} is small compared to that of the condensate part, and 
will be neglected. In accordance with the Thomas-Fermi approximation of the 
condensate, we neglect the kinetic energy of the excited field, too. By moving 
to a frame rotating at the frequency $\mu\rq/\hbar-\omega_{\mathrm{cnt}}$ we 
get a simple, spatially local driving equation
\begin{subequations}
\begin{equation}
\frac{\partial}{\partial t}\delta\hat{\Psi}_{0}(\rr,t)=
-\frac{i}{\hbar}\Delta(\rr)\delta\hat{\Psi}_{0}(\rr,t) 
+\eta(\rr)\hat{I}(t),
\label{eq:diff_eq}
\end{equation}
with a spatially inhomogeneous detuning, 
\begin{equation}
\Delta (\rr)=\frac{1}{\hbar}\left(\hbar\omega_{\mathrm{cnt}}
-\frac{1}{2}\mu_{\mathrm{B}}B_{\mathrm{offs}}
-V_{\mathrm{T}}(\rr)\right) \; ,
\end{equation}
and driving amplitude,
\begin{equation}
\label{eq:eta}
\eta(\rr)=i\sqrt{N}\phi_{\mathrm{BEC}}(\rr) 
\frac{\mu_0 \, \mu_{\mathrm{B}}}{16 \pi  
\sqrt{2}\hbar}\frac{a}{y_0^2}U(\rr) \; .
\end{equation}
\end{subequations}
The time independence of the driving is due to neglecting all terms which 
oscillate with $\omega_{\mathrm{cnt}}$ or $2\omega_{\mathrm{cnt}}$ in the 
rotating frame and average out on time scales longer than 
$1/\omega_{\mathrm{cnt}}$. The magnetic fields $\hat{B}_{i0}$ also average out. 
The dimensionless function $U(\rr)$ expresses the spatial variation of the 
magnetic field modulation due to the CNT,  
\begin{equation}
\label{eq:U_func}
U(\rr)\!=\!\int\limits_{0}^{L}\!\tfrac{x^{2}-2\left(1+y\right)^{2}
+\left(\frac{L}{2}+z-\zeta\right)^{2}-ix\left(1+y\right)}
{\left[x^{2}+\left(1+y\right)^{2}
+\left(\frac{L}{2}+z-\zeta\right)^{2}\right]^{\frac{5}{2}}} 
\sin\left(\!\tfrac{\pi\zeta}{L}\!\right)d\zeta,
\end{equation}
where all the length quantities in the integrand are in units of the CNT-BEC 
distance $y_{0}$. We note that $U(\rr)$ is 
obtained from an infinitely thin 
finite-length current carrying wire that is oscillating as a string clamped at 
both ends.  


Starting with a pure condensate at $t=0$, and letting the system evolve to 
$t=T$  (the measurement time),  the integration of  \ref{eq:diff_eq} 
leads to 
\begin{equation}
\delta\hat{\Psi}_{0}(\rr,T)=
\int_{0}^{T}\eta(\rr)\hat{I}(T-t)e^{-i\Delta(\rr)t}dt \,,
\end{equation}
which expresses the relation between the quantized current $\hat I$ in the CNT 
and the atom field in the magnetic sublevel $m=0$. Atom counting in this 
sublevel allows us to extract quantum statistical properties of the current. We 
assume stationary current, i.e., 
$\left<\hat{I}\left(t'\right)\hat{I}\left(t''\right)\right>= 
\left<\hat{I}\left(0\right)\hat{I}\left(t''-t'\right)\right>$. 
Then the spatially integrated mean number of atoms transferred into the 
sublevel $m=0$ during the measurement time $T$  is 
\begin{multline}
N(\Omega) = \int d^{3}\rr\left<\delta\hat{\Psi}_{0}^{\dagger}(\rr,T)
\delta\hat{\Psi}_{0}(\rr,T)\right>  \\
= T  \int_{-\infty}^{\infty} d\tau \, e^{i\Omega \tau} 
\left<\hat{I}(0)\hat{I}\left(\tau\right)\right> \, f(\tau)\, 
{\mathcal D}(\tau) \; .
\label{eq:N_Omega}
\end{multline} 
The transferred atom number, as being explicitly indicated, is a function of 
the frequency 
$\Omega=\omega_{\mathrm{cnt}}-\tfrac{1}{2}
{\mu_{\mathrm B} B_{\rm offs}}/{\hbar}$, 
which can be finely tuned by the magnetic field $B_\sub{offs}$.  Note that $\omega_{\mathrm{cnt}}$ is typically around 
$2\pi\times 50$ MHz \cite{Sazonova04,Witkamp06}, whereas the  Larmor frequency 
$\frac{1}{2}\mu_{\mathrm{B}}B_{\mathrm{offs}}/\hbar$ can be tuned in the range of 0.1 --- 100 MHz. 

The measurable 
$N(\Omega)$  is related to the current noise spectrum $S(\omega)$ by a 
convolution with the \textit{spectral resolution function}, 
$\mathcal{F}\left\lbrace f(\tau)  {\mathcal D}(\tau)\right\rbrace$,
where $\mathcal{F}\left\lbrace . \right\rbrace$ denotes Fourier transform.
The mapping involves a triangular pulse function,
\begin{equation}
f(\tau)=\left\lbrace \begin{array}{cl}
                      1-\frac{\left|\tau\right|}{T} & \mathrm{if} 
                      \quad \left|\tau\right|\leq T \\
                      0 & \mathrm{else}
                     \end{array}
\right. \; ,
\end{equation}
which originates from the finite measurement time. All properties of the BEC-CNT 
coupling are embedded in 
\begin{equation}
 \label{eq:Dtau}
 {\mathcal D}(\tau) =   \int d^{3}\rr |\eta(\rr)|^2 
 e^{-i \, \tau \, V_{\mathrm T}(\rr)/\hbar}\; .
\end{equation}
Because of the exponential term,  the variation range of the potential energy 
$V_{\mathrm T}(\rr)$ determines the intrinsic bandwidth of the BEC as a probe system (see also 
\ref{fig:d_omega}). This bandwidth is  the chemical potential 
$\mu$, which is  typically in the range of kHz for a  BEC on a chip. 

For very short measurement time $T \ll \hbar/\mu$,  the exponential in the integrand of 
\ref{eq:Dtau} can be approximated by  1, so the spectral resolution is dominated solely by $f(\tau)$.  
On the other hand, for long measurement time $T \gg \hbar/\mu \sim 1$ms, the approximation $f(\tau) = 1-\tfrac{|\tau|}{T} \approx 1$ holds in \ref{eq:N_Omega}
since the function ${\mathcal D}(\tau)$  introduces a cutoff at about 
$\hbar/\mu$. In this limit, the CNT-BEC coupling function ${\mathcal D}(\tau)$  determines the quantum efficiency of the scheme. 

$\mathcal{D}\left(\tau\right)$ can be 
approximated  by neglecting the variation of the magnetic field within the 
condensate, $U\left(\rr\right)\approx U\left(0\right) \equiv U$. This leads to 
the Fourier transform 
\begin{subequations}
\label{eq:resolution_function}
\begin{align}
\label{eq:D_Omega}
\tilde{\mathcal D} (\omega) & \equiv
\mathcal{F}\left\lbrace {\mathcal D}(\tau) \right\rbrace  \approx 
n_{\rm det} \; \tilde{d}  \left(\hbar \omega/ \mu \right),   \\
n_{\rm det} &= \left[\frac{ \mu_0 \, \mu_{\mathrm{B}}}
{16 \pi \sqrt{2} \hbar} \frac{a}{y_0^2}\right]^2 U^{2} N, \\
\tilde{d}(w) &= \left\lbrace \begin{array}{cl}
                \frac{15}{4} \sqrt{w} (1- w) & \mathrm{if} \quad w \leq 1 \\
                0 & \mathrm{else}
                \end{array}
                \right. \; ,
\end{align}
\end{subequations}
where the normalization $\int_{-\infty}^{\infty} dw \, \tilde{d}(w) =1 $  is 
obeyed. In \ref{fig:d_omega}, the approximate $\tilde{\mathcal D}(\omega)$ 
(dotted curve) is compared to exact ones which are obtained numerically for 
different atom numbers $N$ when the CNT length and the CNT-BEC distance is 
fixed. It can be seen that \ref{eq:D_Omega} gives the 
correct order of magnitude and shape of the exact $\tilde{\mathcal D}(\omega)$ for a broad range of the BEC size.

The detectable atom number spectrum, from \ref{eq:D_Omega} is expressed in the form of  the convolution
\begin{equation}
\label{eq:atomspectrum}
N_{\rm long}\left(\tilde\Omega\right) = T \; \frac{\hbar}{\mu} \; n_{\mathrm det}   \int d\tilde\omega \; S(\tilde\omega) \; \tilde{d}(\tilde\Omega-\tilde\omega)\; ,
\end{equation}
where $\tilde\omega$ and $\tilde\Omega$ are in units of $\mu/\hbar$. Measuring the 
atom number at a single value of $\Omega$, the current noise spectrum is readily 
obtained around this frequency with a kHz resolution, i.e., averaged in the bandwidth of the BEC chemical 
potential $\mu/\hbar$. By fine tuning $\Omega$ via the field $B_{\rm offs}$ and using deconvolution, the 
spectrum $S(\omega)$ can be deduced with a much higher resolution.  We recall that $\Omega$ is a frequency relative to the 
CNT vibrational frequency $\omega_{\rm cnt}$. Hence it can be set both positive and negative values, which is substantial to the \textit{quantum galvanometer}.
Finally we note that the maximum frequency range of $S(\omega)$ that can be accessed by the method is limited by  the vibrational frequency $\omega_{\rm cnt}$ to be conform with the rotating wave approximation.


The integral norm of $S(\omega)$  is  $\langle \hat{I}^{2} \rangle$. 
Then, separating the $\Omega$-dependence given by the convolution of normalized 
spectral functions, the coupling strength is on the order of 
$T \frac{\hbar}{\mu} n_{\mathrm det} \langle \hat{I}^{2} \rangle$. For a 
numerical estimate, consider a system described in  
\ref{fig:d_omega}, where an $L=2 \mu$m CNT is oscillating with an amplitude of 
$a=10$ nm \cite{Sazonova04}, at a distance $y_0=4 \mu$m from the BEC, in which 
case $U^{2}\approx 0.4$. For a measurement time $T\approx 1$s that is conform 
with the BEC lifetime in atomchip microtraps, a single atom detected, 
$N_{\rm long}=1$, corresponds to quantum fluctuations of the current on the order of  
$\sqrt{\langle \hat I^2 \rangle} \approx 1$ $\mu$A.

We note that the thermal magnetic near-field noise, which is present in the case of room temperature atomchips due to the trapping wires or metallic coatings on the substrate, has high frequency components resonant with the hyperfine splitting and thus induces spin flips \cite{Henkel1999,Hinds2003,Vuletic2004,Cornell2003}. This parasitic effect must be suppressed by moving the trap to sufficiently large distance from such field sources, while keeping the BEC-CNT distance $y_0$ in the micrometer range, or by using superconducting atomchips \cite{Kasch2010}.

In conclusion, we have evaluated the coupling of trapped atoms to the magnetic field created by the electric current in a mechanically vibrating carbon nanotube. The modeled coupling was found to be strong enough to sense quantum features of the current noise spectrum by means of hyperfine-state-selective atom counting. Hence, our calculations prove that a quantum galvanometer could be realized on the basis of the interaction between a carbon nanotube and a Bose-Einstein condensate of ultracold alkali atoms. Besides the possibility for the experimental realization of a quantum 
galvanometer, the proposed BEC-CNT coupling scheme opens the way to couple 
other degrees of freedom in this hybrid mesoscopic system. For example, one 
could devise `refrigeration' schemes \cite{Zippilli2009} in which heat is 
extracted from the vibrational motion of the CNT and transferred to the 
ultracold gas of atoms. A mechanical/cold atom hybrid quantum system would provide an ideal platform to study thermally driven decoherence mechanisms in nanoscaled quantum systems and would push the sensitivity of mechanical sensors to the ultimate quantum limit. 

This work was supported by the Hungarian National Office for Research and 
Technology under the contract ERC\_HU\_09 OPTOMECH, the Hungarian Academy of Sciences  (Lend\"ulet Program, LP2011-016), and the Hungarian Scientific 
Research Fund (OTKA) under Contract No. K83858. J.~F. acknowledges support by 
the BMBF (NanoFutur 03X5506) and the DFG SFB TRR21.


\begin{thebibliography}{38}
\expandafter\ifx\csname natexlab\endcsname\relax\def\natexlab#1{#1}\fi
\expandafter\ifx\csname bibnamefont\endcsname\relax
  \def\bibnamefont#1{#1}\fi
\expandafter\ifx\csname bibfnamefont\endcsname\relax
  \def\bibfnamefont#1{#1}\fi
\expandafter\ifx\csname citenamefont\endcsname\relax
  \def\citenamefont#1{#1}\fi
\expandafter\ifx\csname url\endcsname\relax
  \def\url#1{\texttt{#1}}\fi
\expandafter\ifx\csname urlprefix\endcsname\relax\def\urlprefix{URL }\fi
\providecommand{\bibinfo}[2]{#2}
\providecommand{\eprint}[2][]{\url{#2}}

\bibitem[{\citenamefont{Harris}(2009)}]{Harris09}
\bibinfo{author}{\bibfnamefont{P.~J.} \bibnamefont{Harris}},
  \textit{\bibinfo{title}{Carbon Nanotube Science}}
  (\bibinfo{publisher}{Cambridge University Press}, \bibinfo{year}{2009}).

\bibitem[{\citenamefont{Kong et~al.}(2000)\citenamefont{Kong, Franklin, Zhou,
  Chapline, Peng, Cho, and Dai}}]{Kong00}
\bibinfo{author}{\bibfnamefont{J.}~\bibnamefont{Kong}},
  \bibinfo{author}{\bibfnamefont{N.~R.} \bibnamefont{Franklin}},
  \bibinfo{author}{\bibfnamefont{C.}~\bibnamefont{Zhou}},
  \bibinfo{author}{\bibfnamefont{M.~G.} \bibnamefont{Chapline}},
  \bibinfo{author}{\bibfnamefont{S.}~\bibnamefont{Peng}},
  \bibinfo{author}{\bibfnamefont{K.}~\bibnamefont{Cho}}, \bibnamefont{and}
  \bibinfo{author}{\bibfnamefont{H.}~\bibnamefont{Dai}},
  \bibinfo{journal}{Science} \textbf{\bibinfo{volume}{287}},
  \bibinfo{pages}{622} (\bibinfo{year}{2000}).

\bibitem[{\citenamefont{Lin et~al.}(2004{\natexlab{a}})\citenamefont{Lin,
  Taylor, Li, Fernando, Qu, Wang, Gu, Zhou, and Sun}}]{Lin04}
\bibinfo{author}{\bibfnamefont{Y.}~\bibnamefont{Lin}},
  \bibinfo{author}{\bibfnamefont{S.}~\bibnamefont{Taylor}},
  \bibinfo{author}{\bibfnamefont{H.}~\bibnamefont{Li}},
  \bibinfo{author}{\bibfnamefont{K.~A.~S.} \bibnamefont{Fernando}},
  \bibinfo{author}{\bibfnamefont{L.}~\bibnamefont{Qu}},
  \bibinfo{author}{\bibfnamefont{W.}~\bibnamefont{Wang}},
  \bibinfo{author}{\bibfnamefont{L.}~\bibnamefont{Gu}},
  \bibinfo{author}{\bibfnamefont{B.}~\bibnamefont{Zhou}}, \bibnamefont{and}
  \bibinfo{author}{\bibfnamefont{Y.-P.} \bibnamefont{Sun}},
  \bibinfo{journal}{J. Mater. Chem.} \textbf{\bibinfo{volume}{14}},
  \bibinfo{pages}{527} (\bibinfo{year}{2004}{\natexlab{a}}).

\bibitem[{\citenamefont{Chiu et~al.}(2008)\citenamefont{Chiu, Hung, Postma, and
  Bockrath}}]{Chiu08}
\bibinfo{author}{\bibfnamefont{H.-Y.} \bibnamefont{Chiu}},
  \bibinfo{author}{\bibfnamefont{P.}~\bibnamefont{Hung}},
  \bibinfo{author}{\bibfnamefont{H.~W.~C.} \bibnamefont{Postma}},
  \bibnamefont{and} \bibinfo{author}{\bibfnamefont{M.}~\bibnamefont{Bockrath}},
  \bibinfo{journal}{Nano Letters} \textbf{\bibinfo{volume}{8}},
  \bibinfo{pages}{4342} (\bibinfo{year}{2008}).

\bibitem[{\citenamefont{H\"{u}ttel et~al.}(2009)\citenamefont{H\"{u}ttel,
  Steele, Witkamp, Poot, Kouwenhoven, and van~der Zant}}]{Huttel09nl}
\bibinfo{author}{\bibfnamefont{A.~K.} \bibnamefont{H\"{u}ttel}},
  \bibinfo{author}{\bibfnamefont{G.~A.} \bibnamefont{Steele}},
  \bibinfo{author}{\bibfnamefont{B.}~\bibnamefont{Witkamp}},
  \bibinfo{author}{\bibfnamefont{M.}~\bibnamefont{Poot}},
  \bibinfo{author}{\bibfnamefont{L.~P.} \bibnamefont{Kouwenhoven}},
  \bibnamefont{and} \bibinfo{author}{\bibfnamefont{H.~S.~J.}
  \bibnamefont{van~der Zant}}, \bibinfo{journal}{Nano Letters}
  \textbf{\bibinfo{volume}{9}}, \bibinfo{pages}{2547} (\bibinfo{year}{2009}).

\bibitem[{\citenamefont{Steele et~al.}(2009)\citenamefont{Steele, Huttel,
  Witkamp, Poot, Meerwaldt, Kouwenhoven, and van~der Zant}}]{Steele09}
\bibinfo{author}{\bibfnamefont{G.~A.} \bibnamefont{Steele}},
  \bibinfo{author}{\bibfnamefont{A.~K.} \bibnamefont{Huttel}},
  \bibinfo{author}{\bibfnamefont{B.}~\bibnamefont{Witkamp}},
  \bibinfo{author}{\bibfnamefont{M.}~\bibnamefont{Poot}},
  \bibinfo{author}{\bibfnamefont{H.~B.} \bibnamefont{Meerwaldt}},
  \bibinfo{author}{\bibfnamefont{L.~P.} \bibnamefont{Kouwenhoven}},
  \bibnamefont{and} \bibinfo{author}{\bibfnamefont{H.~S.~J.}
  \bibnamefont{van~der Zant}}, \bibinfo{journal}{Science}
  \textbf{\bibinfo{volume}{325}}, \bibinfo{pages}{1103} (\bibinfo{year}{2009}).

\bibitem[{\citenamefont{Lassagne et~al.}(2009)\citenamefont{Lassagne,
  Tarakanov, Kinaret, Garcia-Sanchez, and Bachtold}}]{Lassagne09}
\bibinfo{author}{\bibfnamefont{B.}~\bibnamefont{Lassagne}},
  \bibinfo{author}{\bibfnamefont{Y.}~\bibnamefont{Tarakanov}},
  \bibinfo{author}{\bibfnamefont{J.}~\bibnamefont{Kinaret}},
  \bibinfo{author}{\bibfnamefont{D.}~\bibnamefont{Garcia-Sanchez}},
  \bibnamefont{and} \bibinfo{author}{\bibfnamefont{A.}~\bibnamefont{Bachtold}},
  \bibinfo{journal}{Science} \textbf{\bibinfo{volume}{325}},
  \bibinfo{pages}{1107} (\bibinfo{year}{2009}).

\bibitem[{\citenamefont{O'Connell et~al.}(2010)\citenamefont{O'Connell,
  Hofheinz, Ansmann, Bialczak, Lenander, Lucero, Neeley, Sank, Wang, Weides
  et~al.}}]{OConnell2010}
\bibinfo{author}{\bibfnamefont{A.}~\bibnamefont{O'Connell}},
  \bibinfo{author}{\bibfnamefont{M.}~\bibnamefont{Hofheinz}},
  \bibinfo{author}{\bibfnamefont{M.}~\bibnamefont{Ansmann}},
  \bibinfo{author}{\bibfnamefont{R.}~\bibnamefont{Bialczak}},
  \bibinfo{author}{\bibfnamefont{M.}~\bibnamefont{Lenander}},
  \bibinfo{author}{\bibfnamefont{E.}~\bibnamefont{Lucero}},
  \bibinfo{author}{\bibfnamefont{M.}~\bibnamefont{Neeley}},
  \bibinfo{author}{\bibfnamefont{D.}~\bibnamefont{Sank}},
  \bibinfo{author}{\bibfnamefont{H.}~\bibnamefont{Wang}},
  \bibinfo{author}{\bibfnamefont{M.}~\bibnamefont{Weides}},
  \bibnamefont{et~al.}, \bibinfo{journal}{Nature}
  \textbf{\bibinfo{volume}{464}}, \bibinfo{pages}{697} (\bibinfo{year}{2010}).

\bibitem[{\citenamefont{Chu et~al.}(1998)\citenamefont{Chu, Cohen-Tannoudji,
  and Phillips}}]{Nobel1997}
\bibinfo{author}{\bibfnamefont{S.}~\bibnamefont{Chu}},
  \bibinfo{author}{\bibfnamefont{C.~N.} \bibnamefont{Cohen-Tannoudji}},
  \bibnamefont{and} \bibinfo{author}{\bibfnamefont{W.~D.}
  \bibnamefont{Phillips}}, \bibinfo{journal}{Rev. Mod. Phys.}
  \textbf{\bibinfo{volume}{70}}, \bibinfo{pages}{685} (\bibinfo{year}{1998}).

\bibitem[{\citenamefont{Castin}(2001)}]{castin2001}
\bibinfo{author}{\bibfnamefont{Y.}~\bibnamefont{Castin}}, in
  \textit{\bibinfo{booktitle}{'Coherent atomic matter waves', Lecture Notes of
  Les Houches Summer School}}, edited by
  \bibinfo{editor}{\bibfnamefont{R.}~\bibnamefont{Kaiser}},
  \bibinfo{editor}{\bibfnamefont{C.}~\bibnamefont{Westbrook}},
  \bibnamefont{and} \bibinfo{editor}{\bibfnamefont{F.}~\bibnamefont{David}}
  (\bibinfo{publisher}{EDP Sciences and Springer-Verlag},
  \bibinfo{year}{2001}), pp. \bibinfo{pages}{1--136}.

\bibitem[{\citenamefont{Fort\'agh and Zimmermann}(2007)}]{fortagh07}
\bibinfo{author}{\bibfnamefont{J.}~\bibnamefont{Fort\'agh}} \bibnamefont{and}
  \bibinfo{author}{\bibfnamefont{C.}~\bibnamefont{Zimmermann}},
  \bibinfo{journal}{Rev. Mod. Phys.} \textbf{\bibinfo{volume}{79}},
  \bibinfo{pages}{235} (\bibinfo{year}{2007}).

\bibitem[{\citenamefont{Grimm et~al.}(2000)\citenamefont{Grimm,
  Weidem{\"u}ller, and Ovchinnikov}}]{Grimm2000}
\bibinfo{author}{\bibfnamefont{R.}~\bibnamefont{Grimm}},
  \bibinfo{author}{\bibfnamefont{M.}~\bibnamefont{Weidem{\"u}ller}},
  \bibnamefont{and} \bibinfo{author}{\bibfnamefont{Y.~B.}
  \bibnamefont{Ovchinnikov}} (\bibinfo{publisher}{Academic Press},
  \bibinfo{year}{2000}), vol.~\bibinfo{volume}{42} of
  \textit{\bibinfo{series}{Advances In Atomic, Molecular, and Optical Physics}},
  pp. \bibinfo{pages}{95 -- 170}.

\bibitem[{\citenamefont{Wildermuth et~al.}(2005)\citenamefont{Wildermuth,
  Hofferberth, Lesanovsky, Haller, Andersson, Groth, Bar-Joseph, Kr{\"u}ger,
  and Schmiedmayer}}]{Wildermuth2005}
\bibinfo{author}{\bibfnamefont{S.}~\bibnamefont{Wildermuth}},
  \bibinfo{author}{\bibfnamefont{S.}~\bibnamefont{Hofferberth}},
  \bibinfo{author}{\bibfnamefont{I.}~\bibnamefont{Lesanovsky}},
  \bibinfo{author}{\bibfnamefont{E.}~\bibnamefont{Haller}},
  \bibinfo{author}{\bibfnamefont{M.}~\bibnamefont{Andersson}},
  \bibinfo{author}{\bibfnamefont{S.}~\bibnamefont{Groth}},
  \bibinfo{author}{\bibfnamefont{I.}~\bibnamefont{Bar-Joseph}},
  \bibinfo{author}{\bibfnamefont{P.}~\bibnamefont{Kr{\"u}ger}},
  \bibnamefont{and}
  \bibinfo{author}{\bibfnamefont{J.}~\bibnamefont{Schmiedmayer}},
  \bibinfo{journal}{Nature} \textbf{\bibinfo{volume}{435}},
  \bibinfo{pages}{440} (\bibinfo{year}{2005}).

\bibitem[{\citenamefont{Gierling et~al.}(2011)\citenamefont{Gierling,
  Schneeweiss, Visanescu, Federsel, H\"{a}ffner, Kern, Judd, G\"unther, and
  Fort\'{a}gh}}]{Giergling2011}
\bibinfo{author}{\bibfnamefont{M.}~\bibnamefont{Gierling}},
  \bibinfo{author}{\bibfnamefont{P.}~\bibnamefont{Schneeweiss}},
  \bibinfo{author}{\bibfnamefont{G.}~\bibnamefont{Visanescu}},
  \bibinfo{author}{\bibfnamefont{P.}~\bibnamefont{Federsel}},
  \bibinfo{author}{\bibfnamefont{M.}~\bibnamefont{H\"{a}ffner}},
  \bibinfo{author}{\bibfnamefont{D.~P.} \bibnamefont{Kern}},
  \bibinfo{author}{\bibfnamefont{T.~E.} \bibnamefont{Judd}},
  \bibinfo{author}{\bibfnamefont{A.}~\bibnamefont{G\"unther}},
  \bibnamefont{and}
  \bibinfo{author}{\bibfnamefont{J.}~\bibnamefont{Fort\'{a}gh}},
  \bibinfo{journal}{Nature Nanotech.}  (\bibinfo{year}{2011}).

\bibitem[{\citenamefont{Goodsell et~al.}(2010)\citenamefont{Goodsell, Ristroph,
  Golovchenko, and Hau}}]{Goodsell2010Field}
\bibinfo{author}{\bibfnamefont{A.}~\bibnamefont{Goodsell}},
  \bibinfo{author}{\bibfnamefont{T.}~\bibnamefont{Ristroph}},
  \bibinfo{author}{\bibfnamefont{J.~A.} \bibnamefont{Golovchenko}},
  \bibnamefont{and} \bibinfo{author}{\bibfnamefont{L.~V.} \bibnamefont{Hau}},
  \bibinfo{journal}{Phys. Rev. Lett.} \textbf{\bibinfo{volume}{104}},
  \bibinfo{pages}{133002} (\bibinfo{year}{2010}).

\bibitem[{\citenamefont{Petrov et~al.}(2009)\citenamefont{Petrov, Machluf,
  Younis, Macaluso, David, Hadad, Japha, Keil, Joselevich, and
  Folman}}]{Folman09}
\bibinfo{author}{\bibfnamefont{P.~G.} \bibnamefont{Petrov}},
  \bibinfo{author}{\bibfnamefont{S.}~\bibnamefont{Machluf}},
  \bibinfo{author}{\bibfnamefont{S.}~\bibnamefont{Younis}},
  \bibinfo{author}{\bibfnamefont{R.}~\bibnamefont{Macaluso}},
  \bibinfo{author}{\bibfnamefont{T.}~\bibnamefont{David}},
  \bibinfo{author}{\bibfnamefont{B.}~\bibnamefont{Hadad}},
  \bibinfo{author}{\bibfnamefont{Y.}~\bibnamefont{Japha}},
  \bibinfo{author}{\bibfnamefont{M.}~\bibnamefont{Keil}},
  \bibinfo{author}{\bibfnamefont{E.}~\bibnamefont{Joselevich}},
  \bibnamefont{and} \bibinfo{author}{\bibfnamefont{R.}~\bibnamefont{Folman}},
  \bibinfo{journal}{Phys. Rev. A} \textbf{\bibinfo{volume}{79}},
  \bibinfo{pages}{043403} (\bibinfo{year}{2009}).

\bibitem[{\citenamefont{Murphy and Hau}(2009)}]{Murphy09}
\bibinfo{author}{\bibfnamefont{B.}~\bibnamefont{Murphy}} \bibnamefont{and}
  \bibinfo{author}{\bibfnamefont{L.~V.} \bibnamefont{Hau}},
  \bibinfo{journal}{Physical Review Letters} \textbf{\bibinfo{volume}{102}},
  \bibinfo{pages}{033003} (\bibinfo{year}{2009}).

\bibitem[{\citenamefont{Bednorz and
  Belzig}(2010)}]{Bednorz2010Quasiprobabilistic}
\bibinfo{author}{\bibfnamefont{A.}~\bibnamefont{Bednorz}} \bibnamefont{and}
  \bibinfo{author}{\bibfnamefont{W.}~\bibnamefont{Belzig}},
  \bibinfo{journal}{Phys. Rev. Lett.} \textbf{\bibinfo{volume}{105}},
  \bibinfo{pages}{106803} (\bibinfo{year}{2010}).

\bibitem[{\citenamefont{Levitov}(2003)}]{Levitov03}
\bibinfo{author}{\bibfnamefont{L.~S.} \bibnamefont{Levitov}}, in
  \textit{\bibinfo{booktitle}{Quantum Noise In Mesoscopic Physics}}, edited by
  \bibinfo{editor}{\bibfnamefont{Y.~V.} \bibnamefont{Nazarov}}
  (\bibinfo{publisher}{Kluwer Academic Publishers}, \bibinfo{year}{2003}), NATO
  Science Series, II. Math. Phys. and Chem. Vol. 97, pp.
  \bibinfo{pages}{373--396}.

\bibitem[{\citenamefont{Levitov et~al.}(1996)\citenamefont{Levitov, Lee, and
  Lesovik}}]{Levitov96}
\bibinfo{author}{\bibfnamefont{L.~S.} \bibnamefont{Levitov}},
  \bibinfo{author}{\bibfnamefont{H.}~\bibnamefont{Lee}}, \bibnamefont{and}
  \bibinfo{author}{\bibfnamefont{G.~B.} \bibnamefont{Lesovik}},
  \bibinfo{journal}{J. Math. Phys.} \textbf{\bibinfo{volume}{37}},
  \bibinfo{pages}{4845} (\bibinfo{year}{1996}).

\bibitem[{\citenamefont{Vengalattore et~al.}(2007)\citenamefont{Vengalattore,
  Higbie, Leslie, Guzman, Sadler, and Stamper-Kurn}}]{Vengalattore2007High}
\bibinfo{author}{\bibfnamefont{M.}~\bibnamefont{Vengalattore}},
  \bibinfo{author}{\bibfnamefont{J.~M.} \bibnamefont{Higbie}},
  \bibinfo{author}{\bibfnamefont{S.~R.} \bibnamefont{Leslie}},
  \bibinfo{author}{\bibfnamefont{J.}~\bibnamefont{Guzman}},
  \bibinfo{author}{\bibfnamefont{L.~E.} \bibnamefont{Sadler}},
  \bibnamefont{and} \bibinfo{author}{\bibfnamefont{D.~M.}
  \bibnamefont{Stamper-Kurn}}, \bibinfo{journal}{Phys. Rev. Lett.}
  \textbf{\bibinfo{volume}{98}}, \bibinfo{pages}{200801}
  (\bibinfo{year}{2007}).

\bibitem[{\citenamefont{Stibor et~al.}(2010)\citenamefont{Stibor, Bender,
  K\"{u}hnhold, Fort\'{a}gh, Zimmermann, and G\"{u}nther}}]{fortagh_detect}
\bibinfo{author}{\bibfnamefont{A.}~\bibnamefont{Stibor}},
  \bibinfo{author}{\bibfnamefont{H.}~\bibnamefont{Bender}},
  \bibinfo{author}{\bibfnamefont{S.}~\bibnamefont{K\"{u}hnhold}},
  \bibinfo{author}{\bibfnamefont{J.}~\bibnamefont{Fort\'{a}gh}},
  \bibinfo{author}{\bibfnamefont{C.}~\bibnamefont{Zimmermann}},
  \bibnamefont{and}
  \bibinfo{author}{\bibfnamefont{A.}~\bibnamefont{G\"{u}nther}},
  \bibinfo{journal}{New J. Phys.} \textbf{\bibinfo{volume}{12}},
  \bibinfo{pages}{065034} (\bibinfo{year}{2010}).

\bibitem[{\citenamefont{Nazarov and Blanter}(2009)}]{NazarovBlanter}
\bibinfo{author}{\bibfnamefont{Y.~V.} \bibnamefont{Nazarov}} \bibnamefont{and}
  \bibinfo{author}{\bibfnamefont{Y.~M.} \bibnamefont{Blanter}},
  \textit{\bibinfo{title}{Quantum Transport: Introduction to Nanoscience}}
  (\bibinfo{publisher}{Cambridge University Press}, \bibinfo{year}{2009}).

\bibitem[{\citenamefont{Folman et~al.}(2002)\citenamefont{Folman, Kr\"{u}ger,
  Denschlag, Henkel, and Schmiedmayer}}]{Folman_review}
\bibinfo{author}{\bibfnamefont{R.}~\bibnamefont{Folman}},
  \bibinfo{author}{\bibfnamefont{P.}~\bibnamefont{Kr\"{u}ger}},
  \bibinfo{author}{\bibfnamefont{J.}~\bibnamefont{Denschlag}},
  \bibinfo{author}{\bibfnamefont{C.}~\bibnamefont{Henkel}}, \bibnamefont{and}
  \bibinfo{author}{\bibfnamefont{J.}~\bibnamefont{Schmiedmayer}},
  \bibinfo{journal}{Adv. Atom. Mol. Opt. Phys.} \textbf{\bibinfo{volume}{48}},
  \bibinfo{pages}{263} (\bibinfo{year}{2002}).

\bibitem[{\citenamefont{Sazonova et~al.}(2004)\citenamefont{Sazonova, Yaish,
  Ustunel, Roundy, Arias, and McEuen}}]{Sazonova04}
\bibinfo{author}{\bibfnamefont{V.}~\bibnamefont{Sazonova}},
  \bibinfo{author}{\bibfnamefont{Y.}~\bibnamefont{Yaish}},
  \bibinfo{author}{\bibfnamefont{H.}~\bibnamefont{Ustunel}},
  \bibinfo{author}{\bibfnamefont{D.}~\bibnamefont{Roundy}},
  \bibinfo{author}{\bibfnamefont{T.}~\bibnamefont{Arias}}, \bibnamefont{and}
  \bibinfo{author}{\bibfnamefont{P.}~\bibnamefont{McEuen}},
  \bibinfo{journal}{Nature} \textbf{\bibinfo{volume}{431}},
  \bibinfo{pages}{284} (\bibinfo{year}{2004}).

\bibitem[{\citenamefont{Witkamp et~al.}(2006)\citenamefont{Witkamp, Poot, and
  van~der Zant}}]{Witkamp06}
\bibinfo{author}{\bibfnamefont{B.}~\bibnamefont{Witkamp}},
  \bibinfo{author}{\bibfnamefont{M.}~\bibnamefont{Poot}}, \bibnamefont{and}
  \bibinfo{author}{\bibfnamefont{H.~S.~J.} \bibnamefont{van~der Zant}},
  \bibinfo{journal}{Nano Lett.} \textbf{\bibinfo{volume}{6}},
  \bibinfo{pages}{2904} (\bibinfo{year}{2006}).

\bibitem[{\citenamefont{Treutlein et~al.}(2007)\citenamefont{Treutlein, Hunger,
  Camerer, H\"ansch, and Reichel}}]{Treutlein07}
\bibinfo{author}{\bibfnamefont{P.}~\bibnamefont{Treutlein}},
  \bibinfo{author}{\bibfnamefont{D.}~\bibnamefont{Hunger}},
  \bibinfo{author}{\bibfnamefont{S.}~\bibnamefont{Camerer}},
  \bibinfo{author}{\bibfnamefont{T.~W.} \bibnamefont{H\"ansch}},
  \bibnamefont{and} \bibinfo{author}{\bibfnamefont{J.}~\bibnamefont{Reichel}},
  \bibinfo{journal}{Phys. Rev. Lett.} \textbf{\bibinfo{volume}{99}},
  \bibinfo{pages}{140403} (\bibinfo{year}{2007}).

\bibitem[{\citenamefont{Fermani et~al.}(2007)\citenamefont{Fermani, Scheel, and
  Knight}}]{Fermani07}
\bibinfo{author}{\bibfnamefont{R.}~\bibnamefont{Fermani}},
  \bibinfo{author}{\bibfnamefont{S.}~\bibnamefont{Scheel}}, \bibnamefont{and}
  \bibinfo{author}{\bibfnamefont{P.~L.} \bibnamefont{Knight}},
  \bibinfo{journal}{Phys. Rev. A} \textbf{\bibinfo{volume}{75}},
  \bibinfo{pages}{062905} (\bibinfo{year}{2007}).

\bibitem[{\citenamefont{Ristroph et~al.}(2005)\citenamefont{Ristroph, Goodsell,
  Golovchenko, and Hau}}]{Ristroph2005}
\bibinfo{author}{\bibfnamefont{T.}~\bibnamefont{Ristroph}},
  \bibinfo{author}{\bibfnamefont{A.}~\bibnamefont{Goodsell}},
  \bibinfo{author}{\bibfnamefont{J.~A.} \bibnamefont{Golovchenko}},
  \bibnamefont{and} \bibinfo{author}{\bibfnamefont{L.~V.} \bibnamefont{Hau}},
  \bibinfo{journal}{Phys. Rev. Lett.} \textbf{\bibinfo{volume}{94}},
  \bibinfo{pages}{066102} (\bibinfo{year}{2005}).

\bibitem[{\citenamefont{Gr\"uner et~al.}(2009)\citenamefont{Gr\"uner, Jag,
  Stibor, Visanescu, H\"affner, Kern, G\"unther, and Fort\'agh}}]{Fortaghnew}
\bibinfo{author}{\bibfnamefont{B.}~\bibnamefont{Gr\"uner}},
  \bibinfo{author}{\bibfnamefont{M.}~\bibnamefont{Jag}},
  \bibinfo{author}{\bibfnamefont{A.}~\bibnamefont{Stibor}},
  \bibinfo{author}{\bibfnamefont{G.}~\bibnamefont{Visanescu}},
  \bibinfo{author}{\bibfnamefont{M.}~\bibnamefont{H\"affner}},
  \bibinfo{author}{\bibfnamefont{D.}~\bibnamefont{Kern}},
  \bibinfo{author}{\bibfnamefont{A.}~\bibnamefont{G\"unther}},
  \bibnamefont{and}
  \bibinfo{author}{\bibfnamefont{J.}~\bibnamefont{Fort\'agh}},
  \bibinfo{journal}{Phys. Rev. A} \textbf{\bibinfo{volume}{80}},
  \bibinfo{pages}{063422} (\bibinfo{year}{2009}).

\bibitem[{\citenamefont{Steck et~al.}(1998)\citenamefont{Steck, Naraschewski,
  and Wallis}}]{Steck98}
\bibinfo{author}{\bibfnamefont{H.}~\bibnamefont{Steck}},
  \bibinfo{author}{\bibfnamefont{M.}~\bibnamefont{Naraschewski}},
  \bibnamefont{and} \bibinfo{author}{\bibfnamefont{H.}~\bibnamefont{Wallis}},
  \bibinfo{journal}{Phys. Rev. Lett.} \textbf{\bibinfo{volume}{80}},
  \bibinfo{pages}{1} (\bibinfo{year}{1998}).

\bibitem[{\citenamefont{Ho}(1998)}]{Ho1998}
\bibinfo{author}{\bibfnamefont{T.-L.} \bibnamefont{Ho}},
  \bibinfo{journal}{Phys. Rev. Lett.} \textbf{\bibinfo{volume}{81}},
  \bibinfo{pages}{742} (\bibinfo{year}{1998}),
  \urlprefix\url{http://link.aps.org/doi/10.1103/PhysRevLett.81.742}.

\bibitem[{\citenamefont{Henkel et~al.}(1999)\citenamefont{Henkel, P\"{o}tting,
  and Wilkens}}]{Henkel1999}
\bibinfo{author}{\bibfnamefont{C.}~\bibnamefont{Henkel}},
  \bibinfo{author}{\bibfnamefont{S.}~\bibnamefont{P\"{o}tting}},
  \bibnamefont{and} \bibinfo{author}{\bibfnamefont{M.}~\bibnamefont{Wilkens}},
  \bibinfo{journal}{Appl. Phys. B} \textbf{\bibinfo{volume}{69}},
  \bibinfo{pages}{379} (\bibinfo{year}{1999}), ISSN \bibinfo{issn}{0946-2171}.

\bibitem[{\citenamefont{Jones et~al.}(2003)\citenamefont{Jones, Vale, Sahagun,
  Hall, Eberlein, Sauer, Furusawa, Richardson, and Hinds}}]{Hinds2003}
\bibinfo{author}{\bibfnamefont{M.~P.~A.} \bibnamefont{Jones}},
  \bibinfo{author}{\bibfnamefont{C.~J.} \bibnamefont{Vale}},
  \bibinfo{author}{\bibfnamefont{D.}~\bibnamefont{Sahagun}},
  \bibinfo{author}{\bibfnamefont{B.~V.} \bibnamefont{Hall}},
  \bibinfo{author}{\bibfnamefont{C.~C.} \bibnamefont{Eberlein}},
  \bibinfo{author}{\bibfnamefont{B.~E.} \bibnamefont{Sauer}},
  \bibinfo{author}{\bibfnamefont{K.}~\bibnamefont{Furusawa}},
  \bibinfo{author}{\bibfnamefont{D.}~\bibnamefont{Richardson}},
  \bibnamefont{and} \bibinfo{author}{\bibfnamefont{E.~A.} \bibnamefont{Hinds}},
  \bibinfo{journal}{J. Phys. B} \textbf{\bibinfo{volume}{37}},
  \bibinfo{pages}{L15} (\bibinfo{year}{2003}).

\bibitem[{\citenamefont{Lin et~al.}(2004{\natexlab{b}})\citenamefont{Lin,
  Teper, Chin, and Vuleti\ifmmode~\acute{c}\else \'{c}\fi{}}}]{Vuletic2004}
\bibinfo{author}{\bibfnamefont{Y.-j.} \bibnamefont{Lin}},
  \bibinfo{author}{\bibfnamefont{I.}~\bibnamefont{Teper}},
  \bibinfo{author}{\bibfnamefont{C.}~\bibnamefont{Chin}}, \bibnamefont{and}
  \bibinfo{author}{\bibfnamefont{V.}~\bibnamefont{Vuleti\ifmmode~\acute{c}\else
  \'{c}\fi{}}}, \bibinfo{journal}{Phys. Rev. Lett.}
  \textbf{\bibinfo{volume}{92}}, \bibinfo{pages}{050404}
  (\bibinfo{year}{2004}{\natexlab{b}}).

\bibitem[{\citenamefont{Harber et~al.}(2003)\citenamefont{Harber, McGuirk,
  Obrecht, and Cornell}}]{Cornell2003}
\bibinfo{author}{\bibfnamefont{D.~M.} \bibnamefont{Harber}},
  \bibinfo{author}{\bibfnamefont{J.~M.} \bibnamefont{McGuirk}},
  \bibinfo{author}{\bibfnamefont{J.~M.} \bibnamefont{Obrecht}},
  \bibnamefont{and} \bibinfo{author}{\bibfnamefont{E.~A.}
  \bibnamefont{Cornell}}, \bibinfo{journal}{Journal of Low Temperature Physics}
  \textbf{\bibinfo{volume}{133}}, \bibinfo{pages}{229} (\bibinfo{year}{2003}),
  ISSN \bibinfo{issn}{0022-2291}.

\bibitem[{\citenamefont{Kasch et~al.}(2010)\citenamefont{Kasch, Hattermann,
  Cano, Judd, Scheel, Zimmermann, Kleiner, Koelle, and
  Fort\'{a}gh}}]{Kasch2010}
\bibinfo{author}{\bibfnamefont{B.}~\bibnamefont{Kasch}},
  \bibinfo{author}{\bibfnamefont{H.}~\bibnamefont{Hattermann}},
  \bibinfo{author}{\bibfnamefont{D.}~\bibnamefont{Cano}},
  \bibinfo{author}{\bibfnamefont{T.~E.} \bibnamefont{Judd}},
  \bibinfo{author}{\bibfnamefont{S.}~\bibnamefont{Scheel}},
  \bibinfo{author}{\bibfnamefont{C.}~\bibnamefont{Zimmermann}},
  \bibinfo{author}{\bibfnamefont{R.}~\bibnamefont{Kleiner}},
  \bibinfo{author}{\bibfnamefont{D.}~\bibnamefont{Koelle}}, \bibnamefont{and}
  \bibinfo{author}{\bibfnamefont{J.}~\bibnamefont{Fort\'{a}gh}},
  \bibinfo{journal}{New J. Phys.} \textbf{\bibinfo{volume}{12}},
  \bibinfo{pages}{065024} (\bibinfo{year}{2010}).

\bibitem[{\citenamefont{Zippilli et~al.}(2009)\citenamefont{Zippilli, Morigi,
  and Bachtold}}]{Zippilli2009}
\bibinfo{author}{\bibfnamefont{S.}~\bibnamefont{Zippilli}},
  \bibinfo{author}{\bibfnamefont{G.}~\bibnamefont{Morigi}}, \bibnamefont{and}
  \bibinfo{author}{\bibfnamefont{A.}~\bibnamefont{Bachtold}},
  \bibinfo{journal}{Phys. Rev. Lett.} \textbf{\bibinfo{volume}{102}},
  \bibinfo{pages}{096804} (\bibinfo{year}{2009}).

\end{thebibliography}

\end{document}